\documentclass{article}
\usepackage{frascatiphys
}
\usepackage{amssymb}
\usepackage{amsmath}
\usepackage{amsfonts}
\usepackage[]{hyperref}
\begin{document}
\title{ 
INCLUDING QCD RADIATION CORRECTIONS \\
IN TRANSPLANCKIAN SCATTERING
}
\author{
Paolo Lodone        \\
{\em Scuola Normale Superiore and INFN, Pisa, Italy} \\
}
\maketitle
\baselineskip=11.6pt
\begin{abstract}
In the context of Large Extra Dimensions the fundamental Planck scale can be as low as the TeV scale.
If this is the way of nature, quantum gravity effects could be visible at the LHC or other High energy colliders.
A model independent signal is given by elastic scattering at small angles and transplanckian energy, which can be computed in the eikonal approximation.
It is interesting both from a phenomenological and a theoretical point of view to study the corrections to this process due to the QCD radiation.
To include it, we generalize a shock-wave method proposed by 't Hooft, so that we are able to obtain the amplitude at first order in the gauge interaction but resummed to all orders in gravity. From this result we can learn many interesting things, for example we can extract the true scale of the process which agrees with an earlier guess of Emparan, Masip and Rattazzi. \\
The problem of radiation in transplanckian scattering is briefly introduced, we then report and comment the recent results and mention the perspectives for the near future.  
\end{abstract}
\baselineskip=14pt

\section{Introduction}

Motivated by the Hierarchy problem, the hypothesis of Large Extra Dimensions \cite{arkanihamed} (LED) is that there may exist $n$ compactified extra dimensions so that the Einstein - Hilbert action of General Relativity (GR) becomes:
\begin{equation}
S_{D} = \frac{1}{2} \int d^{D}x \sqrt{-g^{(D)}} M_D^{n+2} R^{(D)}.
\end{equation}
where $D=4+n$. Integrating out these extra dimensions one obtains:
\begin{equation}
M_{Pl}^2 = M_D^{2+n} (2\pi r)^n
\end{equation}
where $r$ is the size of the extra dimensions in the simplest case of toroidal compactification.
This means that if the volume of the compactified extra dimensions is large in TeV units, then the scale $M_D$ at which Quantum Gravity (QG) should manifest itself is much lower then the usual Planck scale $M_{Pl}$.

The important point (see \cite{Kribs:2006mq}\cite{Csaki:2004ay} for a review) is that, if we assume that only gravity can propagate in the $n$ extra dimensions, then observable effects are to be expected only in High Energy Physics and in deviations from the newtonian potential at distances smaller than $r$. For example setting $M_D = 1$ TeV we have $r = 2 \cdot 10^{-16} 10^{\frac{32}{n}} \mbox{ mm}$.
Since the experimental data impose $r \leq 0.2$ mm, we immediatly see that:
\begin{equation}
n \geq 2 \qquad , \qquad M_D \approx 1 \mbox{ TeV}
\end{equation}
is a serious possibility.
On the other hand in high energy physics the processes of phenomenological interest are elastic scattering, for example $p-p$ at the LHC \cite{giudice1} or $\nu - p$ for the physics of cosmic rays \cite{Emparan:2001kf}, and black hole creation, which however is model dependent (see \cite{Rattazzi:2002rz} for a brief overall review). We will consider elastic scattering, which can be nicely computed in the eikonal regime, leaving out black hole creation over which there are serious theoretical uncertainties.

\section{Transplanckian scattering}

Defining the $D$-dimensional Newton constant as $G_D = \frac{(2\pi)^{n-1} \hbar^{n+1}}{4 c^{n-1} M_D^{n+2}}$, the relevant length scales are:
\begin{equation}
\lambda_B = \frac{4\pi \hbar c}{\sqrt{s}} \quad , \quad \lambda_P = \left( \frac{G_D \hbar}{c^3} \right)^{\frac{1}{n+2}} \quad , \quad R_S = \frac{1}{\sqrt{\pi}} \left[ \frac{8 \Gamma (\frac{n+3}{2})}{n+2}  \right]^{\frac{1}{n+1}}
\end{equation}
where $\lambda_B$ is the de Broglie wavelength, $\lambda_P$ is the Planck scale (at which QG appears), and $R_S$ is the Schwarzschild radius (at which curvature effects become large).
It is important to notice that in this $D$-dimensional framework we also have the length scale:
\begin{equation}
b_c = \left( \frac{G_D s}{\hbar c^5} \right)^{\frac{1}{n}}
\end{equation}
which can not be defined if $n=0$ and moreover it goes to infinity if $\hbar \rightarrow 0$ with $G_D$ fixed. We can thus say that $b_c$ is related to the size of the classical region in the impact parameter space.

Since the true theory of QG is not known, we are interested in model independent predictions.
First of all let us assume that $b_c \gg R_S$, which is certainly true at sufficiently high energy.
Notice that for impact parameter $b \gg R_S$ gravity can be linearized. 
Moreover if $\sqrt{s} \gg M_D$ then $R_S \gg \lambda_P \gg \lambda_B$, which means that QG effects are expected to be small. Finally, in the case of forward scattering at small angles, we are able to perform a predictive computation using the eikonal resummation or the shock-wave method, as we will see in the following. In conclusion, in the transplanckian eikonal regime defined by:
\begin{equation} \label{transeik}
\sqrt{s} \gg M_D \qquad , \qquad  \frac{-t}{s}\ll 1
\end{equation}
it is possible to obtain a neat and model independent prediction which relies only on Quantum Mechanics (QM) and linearized GR.
Notice that all this can be of phenomenological interest only in a LED scenario with the QG scale $M_D$ lowered down to a few TeV.

\section{Eikonal amplitude without radiation}

As shown in \cite{giudice1} and \cite{kabat}, a first approach for the evaluation of the transplanckian amplitude at small angles is the so-called  ``eikonalization'', which amounts to resum an infinite number of Feynman diagrams with graviton exchanges which are one by one ultraviolet (UV) divergent but whose sum is finite and independent of the regulator. In performing this sum one neglects the virtuality of gravitons in the matter propagators and makes use of the on-shell vertices.
The final result is:
\begin{equation} \label{eikonalamplitude}
\mathcal{A}_{eik}(q_{\perp}) = -2is \int d^2 b_{\perp} e^{-i q_{\perp} b_{\perp}} (e^{i \chi(b_{\perp})} -1)
\end{equation}
where $q_{\perp}$ is the momentum transfer, which is mainly transverse ($q_{\perp}^2 \approx -t$), and:
\begin{equation} \label{eikonal}
\chi(b_{\perp}) = \frac{1}{2s} \int \frac{d^2 q_\perp}{(2\pi)^2} e^{iq_{\perp} b_{\perp}} \mathcal{A}_{tree}(q_{\perp}) = \left( \frac{b_c}{b_{\perp}} \right)^n .
\end{equation}
Notice that this amplitude is spin independent and moreover there is no UV sensitivity. This means that in this regime we can obtain a prediction in perturbative QG which does not depend on the details of the UV completion.

However, as we said, this prediction relies only on QM and GR, and thus it should be computable without passing through Feynman diagrams. This is realized in a different approach which is due to 't Hooft \cite{thooft}. The idea is to replace one of the two particles with its gravitational field, and then to solve the Equations Of Motion (EOM) for the other particle in this background.
The solution of the Einstein equations for a very energetic particle is the Aichelburg-Sexl Shock Wave (SW) metric \cite{aichelburg}:
\begin{equation}
d s^2 = - d x^{+} dx^{-} + \Phi(x_{\perp}) \delta(x^{-}) (dx^{-})^2 + dx_{\perp}^2
\end{equation}
where the particle is moving in the positive $z$ direction, while:
\begin{eqnarray}
\frac{\Phi}{8\pi G_{D}} & =& -\frac{E}{\pi}\log|x_{\perp}| \qquad \qquad \qquad \mbox{ if } D=4 \nonumber \\
& =& \frac{2\, \Gamma(\frac{k+1}{2})\, E}{2\, \pi^{\frac{k+1}{2}}\, (D-4)\, |x_{\perp}|^{D-4}} \qquad \mbox{ if } D>4 \, .
\end{eqnarray}
and we have generalized the SW to the case of $n=D-4$ extra dimensions.
All one has to do is to solve the Klein Gordon equation in this background (we consider a scalar particle for simplicity).
A fast way to do this \cite{rychkov1} is to perform a discontinuous coordinate transformation $x \rightarrow x^{\prime}$ in order to make the metric continuous across $x^{-}=0$. At $x^{-}=x^{\prime-} =0$ we have:
\begin{equation} \label{coordinatetransformation}
\left\{
\begin{array}
[c]{l}%
x^{+}=x^{\prime+}+\theta(x^{-})\, \Phi(x_{\perp}) \\
x^{i}=x^{\prime i} \qquad (i=1,2)\, \, .
\end{array}
\right.
\end{equation}
Then we solve the equations of motion for the other particle in this shock-wave spacetime, which consists of two flat semispaces glued together at $x^{-}=0$ with the discontinuity (\ref{coordinatetransformation}). For a particle with energy $E$ which is moving in the negative $z$ direction ($p^- = 2E, \, p^+ = p_\perp =0$) the wavefuction is, before the collision:
\begin{equation}
\psi(x) = e^{-i p x} = e^{-iE x^{+}} \qquad (x^{-} < 0)
\end{equation}
while immediatly after the collision the continuity in the $x^{\prime}$ coordinates implies:
\begin{equation}
\psi(x) = e^{-iE(x^{+} - \Phi(x_{\perp}))} \qquad (x^{-} \rightarrow 0^{+})
\end{equation}
thus we obtain the eikonal amplitude in inpact parameter representation:
\begin{equation}
\mathcal{A}_{eik}(x_{\perp}) = e^{i E \Phi(x_{\perp})}.
\end{equation}
It is easy to see that $E \Phi(x_{\perp})$ is exactly the eikonal $\chi(x_{\perp})$ of equation (\ref{eikonal}).
Notice that the $-1$ term in (\ref{eikonalamplitude}) gives a contribution proportional to $\delta(q_{\perp})$, which corresponds to the trivial part of the scattering, while the overall factor can be obtained by properly normalizing the wavefunctions.

Notice that the total elastic cross section is given by (see \cite{giudice1}):
\begin{equation}
\sigma_{el} = \frac{\mbox{Im}\,  \mathcal{A}_{eik} (0)}{s} = 2 \pi b_c^2 \Gamma (1-\frac{2}{n}) \cos \frac{\pi}{n} \quad \sim \quad s^{2/n}
\end{equation}
which is basically the area of a disk of radius $b_c$ (apart from the case $n=2$), while that for black hole creation can be only estimated as:
\begin{equation}
\sigma_{bh} \approx \pi R_S^2 \quad \sim \quad s^{1/(n+1)}
\end{equation}
which is parametrically smaller.

\section{The problem of radiation}
Besides the theoretical interest in the subject \textit{per se}, why should one worry about attaching radiation to the eikonal amplitude (\ref{eikonalamplitude})? From a phenomenological point of wiew we are interested in the case in which at least one of the two incoming particles is a hadron. In our case the momentum transfer $q_\perp$ is typically much larger than the QCD scale, for example for values of $M_D$ which are relevant at the at the LHC we have $b_c \sim (100\mbox{ GeV})^{-1}$. This means that the typical impact parameter is much smaller then the proton size, and we expect the parton picture to be applicable (ie no multiple parton interaction). We then write:
\begin{equation}
\frac{d \sigma_{tot}}{dt} = \int_0^1 dx \, \sum_j f_j(x, \mu_f) \, \frac{d \sigma_j}{dt}
\end{equation}
where $f_j$ are the Parton Distribution Functions (PDF) and $\mu_f$ is the factorization scale at which they are normalized. This scale is choosen in order to minimize the higher order corrections to the cross section, for example in the usual Deep Inelastic Scattering (DIS) the correct prescription is $\mu_f = q$. This scaling violating evolution of the PDF is actually the main effect which is due to the QCD radiation, and it is taken into account by just normalizing the PDF at the correct scale (see eg \cite{QCD} for a full reference).

Thus the problem of including radiation amounts first of all to determine the correct factorization scale.
The reason why we expect something nonstandard is very intuitive.
The usual DIS is a pretty quantum phenomenon in which the momentum transfer $q$ is basically carried by a single particle, for example a photon $\gamma^*$. In Transplanckian scattering instead the same is true only if the impact parameter is larger than $b_c$, ie $q < b_c^{-1}$. If $q> b_c^{-1}$ then the process is semiclassical, which means that a large number of graviton is exchanged, with the result that the typical momentum of an exchanged graviton is now $q$ divided by the total number of gravitons which are exchanged. For this heuristic reason, in \cite{Emparan:2001kf} the following factorization scale was proposed:
\begin{equation}
\mu_{f}(q)=\left\{
\begin{array}
[c]{l}%
q\,\quad\text{if}\quad q<b_{c}^{-1}\,,\\
q^{\frac{1}{n+1}}\left( b_{c}^{-1}\right) ^{\frac{n}{n+1}}\,\text{\quad
if\quad}q>b_{c}^{-1}\,.
\end{array}
\right. \label{eq:mu}%
\end{equation}
The first purpose in including radiation in a consistent way is to check this nonstandard fact through a direct computation.

\section{Including radiation}

In order to compute the QCD radiation corrections to the amplitude (\ref{eikonalamplitude}) the best way to proceed is to generalize 't~Hooft method, in which the amplitude emerges as a whole in a nonperturbative way.
To do this, we consider the simplified case in which only one of the two incoming particles can radiate.
We then replace the other particle with its gravitational field and do perturbative Quantum Field Theory  in the SW background for the remaining particle \textit{and} its radiation.
This is what is done in \cite{Lodone:2009qe}in full detail: the relevant Feynman rules are there derived together with a formalism which allows to easily write down the relativistic amplitudes with the two particles in the initial state.

In this section the same result for the amplitude with radiation will be obtained in a more direct way, which is basically perturbation theory in position space, and which makes even more evident the connection with 't Hooft method. Rewrite the eikonal amplitude as:
\begin{equation}
I(p^{-},q_\perp) \equiv\int d^{2} x_\perp\,e^{-i q_\perp x_\perp}e^{i\frac{1}{2}p^{-}\Phi(x_\perp)}\,.
\end{equation}
Then the solutions of the EOM which reduce to plane waves in the past or in the future are:
\begin{eqnarray}
\phi_{p}^{\text{in}}(x) &=& \theta(-x^{-})e^{i[p].x}%
+\theta(x^{-})\int\frac{d^{2} q_\perp}{(2\pi)^{2}}\,I(p^{-},q_\perp%
)\,e^{i[p+ q_\perp].x}\,, \\
\phi_{p}^{\text{out}}(x) &=& \theta(x^{-})e^{i[p].x}%
+\theta(-x^{-})\int\frac{d^{2} q_\perp}{(2\pi)^{2}}\,I(p^{-},q_\perp%
)\,e^{i[p- q_\perp].x}\, ,
\end{eqnarray}
where the \textquotedblleft vector in square brackets\textquotedblright \ notation denotes an on-shell 4-vector whose $+$ component is computed in terms of the known $-$ and $\mathbf{\perp}$ components, i.e. $[p+ q_\perp]^+ \equiv (p_\perp+q_\perp)^{2}/p^{-}$.
Analogously one has to solve the EOM for a vector particle, obtaining $A_{l,\varepsilon}^{\text{in-out}}(x)_\mu^a$, and one has to take care of the discontinuity of the polarization across the SW.
The simplest thing to do is to use the covariant light cone gauge:
\begin{equation}
l^\mu \varepsilon_\mu =0 \quad , \quad \varepsilon_+ =0
\end{equation}
so that one can fix $\varepsilon_-$ in terms of $l$ and $\varepsilon_\perp$ and forget about the change in polarization across the SW, since $\varepsilon_\perp$ is continuous (we refer to \cite{Lodone:2009qe} for details).

Consider now for simplicity a scalar quark with momentum $p$ which scatters across the SW changing its momentum into $p^{\prime}$ and emitting a gluon with momentum $l$.
Then apart from some overall factors the matrix element is:
\begin{equation}
\mathcal{M}_{p\rightarrow p^{\prime}+l} \propto \int d^{4}x\sqrt{-g}g^{\mu\nu}\phi_{p'}^{\text{out}}(x) \overleftrightarrow{\partial}_{\mu}\phi_{p}^{\text{in}}(x) \, A_{l,\varepsilon}^{\text{out}}(x)_\nu^a .%
\end{equation}
This integral can be decomposed in the cases in which the gluon is emitted  respectively after and before the SW, ie $\mathcal{M} = \int_{x^{-}>0} + \int_{x^{-}<0} = \mathcal{M}^{(\text{I)}} + \mathcal{M}^{(\text{II)}} $. Apart from normalizations and delta functions one obtains:
\begin{align}
\mathcal{M}^{(\text{I)}} & \propto + g_{s}T_{ij}^{a}\, \, 2 p^{\prime -} I(p^{-},q_\perp)
 \, \frac{(l_\perp - \frac{l^-}{p^-} q_\perp)\varepsilon_\perp}{(l_\perp - \frac{l^-}{p^-} q_\perp)^2 } \,, \label{eq:M1geval:initial} \\
\mathcal{M}^{(\text{II})} & \propto - g_{s}T_{ij}^{a} \, \, 2p^- \int\frac{d^{2}k_\perp%
}{(2\pi)^{2}}  I(l^{-},k_\perp)\,I(p^{\prime-},q_\perp-k_\perp%
) \frac{(l_\perp - k_\perp)\varepsilon_\perp}{(l_\perp - k_\perp)^2}\,. \label{eq:M1geval}%
\end{align}
This is the correct amplitude at lowest order in QCD but nonperturbative in gravity, in the transplanckian regime.
Notice from (\ref{eq:M1geval}) that the gluon re-interacts with the SW if it is emitted before the collision. Notice also that in the soft limit $l^- \rightarrow 0$ one has $I(l^{-},k_\perp) \rightarrow (2\pi)^2 \, \delta (k_\perp)$, so that the integral in $k_\perp$ becomes trivial and the sum of $\mathcal{M}^{(\text{I)}}$ and $\mathcal{M}^{(\text{II)}}$ becomes equal to the amplitude without radiation times the Weinberg current:
\begin{equation}
\left( \frac{p_1^\mu}{p_1 l} - \frac{p_1^{\prime \mu}}{p^\prime_1 l} \right)\varepsilon_\mu = -2 \left( \frac{l_\perp}{(l_\perp)^2} - \frac{p^{\prime -}}{p^-} \, \frac{l_\perp - \frac{l^-}{p^-} q_\perp}{(l_\perp - \frac{l^-}{p^-} q_\perp)^2} \right) \varepsilon_\perp \, .
\end{equation}
This corresponds to the fact that the gravitational interaction of the gluon is negligible if its energy is small.

As a first application of this result, we can determine the scale of the process by studying the large logarithms coming from $\int {d^2 l} \left| \mathcal{M}_{p\rightarrow p^{\prime}+l} \right|^2$.
We focus on the initial state radiation (\ref{eq:M1geval}), which is directly related to the evolution of the PDF (while the final state radiation (\ref{eq:M1geval:initial}) is clearly completely standard).
To this purpose, the relevant enhancement is not that related to softness, instead it is that related to collinearity $l_\perp \ll q_\perp$, while $l^- / p^-$ can be a relevant fraction of the total energy.
Remember that, in general, factorization means the following fact (see \cite{Nason} for a practical review). Consider an amplitude $M_{m+1}$ in which there are $m+1$ external final state legs, one of which is a photon with momentum $l$. Then in the limit in which $l$ becomes collinear to one of the other legs (call $x$ this leg and $l_\perp$ the perpendicular component with respect to the direction of this leg) we have:
\begin{equation} \label{factorization}
|M_{m+1}|^2 d \Phi_{m+1}  \rightarrow |M_m|^2 d \Phi_m \, \frac{\alpha_S}{2\pi} P_{x\rightarrow xg}(z) dz \, \frac{d\phi}{2\pi} \, \frac{d |l_\perp|}{|l_\perp|}
\end{equation}
where $P_{x \rightarrow xg}$ is the splitting function for particle $x$ in $x$ plus gluon, and $z$ is the energy fraction which remains to particle $x$ after the splitting. For a radiated gluon, one has to integrate from an infrared cutoff $\mu_{IR}$ which is basically the QCD scale up to $l_\perp$ of order of the momentum transfert $q$. This is the reason why (see \cite{QCD} for a full account) one gets large logarithms $\log \frac{q}{\mu_{IR}}$ which can be reabsorbed into the PDF once they are evolved according to the Dokshitzer-Gribov-Lipatov-Altarelli-Parisi (DGLAP) equations and normalized at the scale $q$.
Let us see what happens in the transplanckian case. It can be shown \cite{Lodone:2009qe} that, if $l^-$ is a relevant fraction of $p^-$, then instead of reproducing (\ref{factorization}) the modulus square of (\ref{eq:M1geval}), multiplied by the phase space of the gluon, behaves as:
\begin{equation} \label{eq:gluondistribution}
\frac{d |l_\perp|}{|l_\perp|} \quad \rightarrow \quad 
\begin{array}{ll}
\frac{d|l_\perp|}{|l_\perp|} & \mbox{ if } |l_\perp| \ll b_c^{-1} \\
\frac{1}{n+1} \frac{d|l_\perp|}{|l_\perp|} & \mbox{ if } |l_\perp| \gg b_c^{-1} 
\end{array} \, .
\end{equation}
Integrating over this peculiar gluon distribution up to $l_\perp = q_\perp$ one obtains exactly the factorization scale (\ref{eq:mu}), thus confirming the proposal of \cite{Emparan:2001kf} through an explicit computation.

\section{Conclusions and perspectives}
The problem of taking radiation into account in transplanckian elastic scattering has been shown to become a tractable problem after reducing it to a computation of Quantum Field Theory in curved background. This can be done by generalizing an earlier idea of 't Hooft, in which one of the two particles is replaced by its gravitational field, the Aichelburg-Sexl Shock Wave.

Using this method, it has been proved the proposal of Emparan, Masip and Rattazzi about the correct factorization scale in transplanckian processes involving hadrons. 
In doing so, one discovers that the probability distribution of the radiated partons has a peculiar shape, which depends directly on the number of extra dimensions. Thus, related to this fact, there could be interesting experimental signatures at colliders, which require further investigation.

\section{Acknowledgements}

In this project I have worked jointly with Vyacheslav Rychkov, whom I thank.

\end{document}